\documentclass[10pt,twoside]{article}
\usepackage{amssymb}
\def\firstpage{1}\def\lastpage{1000}

\makeatletter
\def\magnification{\afterassignment\m@g\count@}
\def\m@g{\mag=\count@\hsize6.5truein\vsize8.9truein\dimen\footins8truein}
\makeatother

\font\caps=cmcsc10                    % Theorem, Lemma etc
\font\Caps=cccsc10 scaled \magstep1   % Title
\font\scaps=cmcsc8

\makeatletter
\@specialpagefalse\headheight=8.5pt
\def\DocMath{{\def\th{\thinspace}\scaps Doc.\th Math.\th J.\th DMV}}
\renewcommand{\@oddfoot}{\hfill\scaps Documenta Mathematica $\cdot$ Extra
        Volume ICM  1998 $\cdot$ \number\firstpage--\lastpage\hfill}
\renewcommand{\@evenfoot}{\ifnum\thepage>\lastpage\hfill\scaps
    Documenta Mathematica $\cdot$ Extra Volume ICM  1998 $\cdot$
\hfill\else\@oddfoot\fi}%
\renewcommand{\@evenhead}{%
    \ifnum\thepage>\lastpage\rlap{\thepage}\hfill%
    \else\rlap{\thepage}\slshape\leftmark\hfill\caps\SAuthor\hfill\fi}%
\renewcommand{\@oddhead}{%
    \ifnum\thepage=\firstpage{\DocMath\hfill\llap{\thepage}}%
    \else{\slshape\rightmark}\hfill\caps\STitle\hfill\llap{\thepage}\fi}%
\makeatother

\def\TSkip{\bigskip}
\newbox\TheTitle{\obeylines\gdef\GetTitle #1
\ShortTitle  #2
\SubTitle    #3
\Author      #4
\ShortAuthor #5
\EndTitle
{\setbox\TheTitle=\vbox{\baselineskip=20pt\let\par=\cr\obeylines%
\halign{\centerline{\Caps##}\cr\noalign{\medskip}\cr#1\cr}}%
	\copy\TheTitle\TSkip\TSkip%
\def\next{#2}\ifx\next\empty\gdef\STitle{#1}\else\gdef\STitle{#2}\fi%
\def\next{#3}\ifx\next\empty%
    \else\setbox\TheTitle=\vbox{\baselineskip=20pt\let\par=\cr\obeylines%
    \halign{\centerline{\caps##} #3\cr}}\copy\TheTitle\TSkip\TSkip\fi%
\centerline{\caps #4}\TSkip\TSkip%
\def\next{#5}\ifx\next\empty\gdef\SAuthor{#4}\else\gdef\SAuthor{#5}\fi%
\catcode'015=5}}\def\Title{\obeylines\GetTitle}
\def\Abstract{\begingroup\narrower
    \parskip=\medskipamount\parindent=0pt{\caps Abstract. }}
\def\EndAbstract{\par\endgroup\TSkip}

\long\def\MSC#1\EndMSC{\def\arg{#1}\ifx\arg\empty\relax\else
     {\par\narrower\noindent%
     1991 Mathematics Subject Classification: #1\par}\fi}

\long\def\KEY#1\EndKEY{\def\arg{#1}\ifx\arg\empty\relax\else
	{\par\narrower\noindent Keywords and Phrases: #1\par}\fi\TSkip}

\newbox\TheAdd\def\Addresses{\vfill\copy\TheAdd\vfill
    \ifodd\number\lastpage\vfill\eject\phantom{.}\vfill\eject\fi}
{\obeylines\gdef\GetAddress #1
\Address #2
\Address #3
\Address #4
\EndAddress
{\def\xs{6truecm}
\setbox0=\vtop{{\obeylines\hsize=\xs#1\par}}\def\next{#2}
\ifx\next\empty % 1 address
     \setbox\TheAdd=\hbox to\hsize{\hfill\copy0\hfill}
\else\setbox1=\vtop{{\obeylines\hsize=\xs#2\par}}\def\next{#3}
\ifx\next\empty % 2 addresses
     \setbox\TheAdd=\hbox to\hsize{\hfill\copy0\hfill\copy1\hfill}
\else\setbox2=\vtop{{\obeylines\hsize=\xs#3\par}}\def\next{#4}
\ifx\next\empty\ % 3 addresses
     \setbox\TheAdd=\vtop{\hbox to\hsize{\hfill\copy0\hfill\copy1\hfill}
                \vskip20pt\hbox to\hsize{\hfill\copy2\hfill}}
\else\setbox3=\vtop{{\obeylines\hsize=\xs#4\par}}
     \setbox\TheAdd=\vtop{\hbox to\hsize{\hfill\copy0\hfill\copy1\hfill}
	        \vskip20pt\hbox to\hsize{\hfill\copy2\hfill\copy3\hfill}}
\fi\fi\fi\catcode'015=5}}\gdef\Address{\obeylines\GetAddress}

\begin{document}

%%%%% ------------- fill in your data below this line  -------------------
%%%%%    The following lines \Title ... \EndAddress must ALL be present
%%%%%    and in the given order.
\Title
%%%%%    Put here the title. Line breaks will be recognized. 
The Mathematics of Fivebranes
\ShortTitle
%%%%%    Running title for odd numbered pages, ONE line, please. 
%%%%%    If none is given, \Title will be used instead.          
\SubTitle
%%%%%    A possible subtitle goes here.
\Author
%%%%%    Put here name(s) of authors. Line breaks will be recognized.  
Robbert Dijkgraaf
\ShortAuthor
%%%%%%   Running title for even numbered pages, ONE line, please. 
%%%%%%   If none is given, \Author will be used instead.          
\EndTitle
\Abstract
%%%%%    Put here the abstract of your manuscript.

Fivebranes are non-perturbative objects in string theory that generalize
two-dimensional conformal field theory and relate such diverse subjects
as moduli spaces of vector bundles on surfaces, automorphic forms,
elliptic genera, the geometry of Calabi-Yau threefolds, and generalized
Kac-Moody algebras.

\EndAbstract
\MSC
%%%%%    1991 Mathematics Subject Classification: 
81T30
\EndMSC
\KEY
%%%%%    Keywords and Phrases:     
quantum field theory,  elliptic genera, automorphic forms
\EndKEY
%%%%%    All 4 \Address lines below must be present. To center the last
%%%%%    entry, no empty lines must be between the following \Address
%%%%%    and \EndAddress lines.
\Address
%%%%%    Address of first Author here
Robbert Dijkgraaf
Department of Mathematics
University of Amsterdam
Plantage Muidergracht 24
1018 TV Amsterdam
The Netherlands
rhd@wins.uva.nl
\Address
%%%%%    Address of second Author here etc.
\Address
\Address
\EndAddress
%%
%%       Make sure the last tex command in your manuscript
%%       before the first \end{document} is the command  \Addresses
%%
%%---------------------Here the prologue ends---------------------------------
%%--------------------Here the manuscript starts------------------------------

\newcommand{\twomatrix}[4]{{\left(\begin{array}{cc}#1 & #2\\
#3 & #4 \end{array}\right)}}
\hyphenation{di-men-sion-al}                       
\hyphenation{di-men-sion-al-ly}
\newcommand{\ie}{{\it i.e.}}
\newcommand{\etc}{{\it etc.\ }}
\newcommand{\cf}{{\it cf}\ }
\newcommand{\eg}{{\it e.g.\ }}
\newcommand{\smallfrac}[2]{{\textstyle #1 \over #2}}
\newcommand{\1}{{\it 1}}
\newcommand{\Z}{{\Bbb Z}}
\newcommand{\R}{{\Bbb R}}
\newcommand{\C}{{\Bbb C}} 
\renewcommand{\P}{{\Bbb P}}
\renewcommand{\H}{{\Bbb H}}
\newcommand{\cH}{{\cal H }} 
\newcommand{\cL}{{\cal L }}            
\newcommand{\zbar}{{\overline{z}}}
\newcommand{\Lbar}{{\overline{L}}}
\newcommand{\Tr}{{\rm Tr}\,}
\newcommand{\taubar}{{\overline{\tau}}}
\newcommand{\cM}{{\cal M}}
\newcommand{\del}{\partial}
\def\a{\alpha} 
\def\b{\beta} 
\def\g{\gamma} 
\def\G{\Gamma}
\def\e{\epsilon}
\def\h{\eta}
\def\th{\theta} 
\def\Th{\Theta}  
\def\k{\kappa}
\def\l{\lambda} 
\def\L{\Lambda} 
\def\m{\mu}
\def\n{\nu}
\def\r{\rho} 
\def\s{\sigma} 
\def\t{\tau}
\def\f{\phi} 
\def\F{\Phi} 
\def\w{\omega}
\def\W{\Omega} 
\def\v{\varphi} 
\def\z{\zeta}
\def\<{\langle}
\def\>{\rangle}

\newcommand{\SNX}{{S^N\!X}}
\newcommand{\HNX}{{\rm Hilb}^N\!(X)}

\newcommand{\id}{{\bf 1}}

% figures
%
\input epsf
\newcommand{\insertfig}[2]{\leavevmode \vcenter{\hbox{\epsfxsize=#2 cm
\epsffile{#1.eps}}}}
\newcommand{\be}{\begin{equation}}
\newcommand{\ee}{\end{equation}}

\section{Introduction}

This joint session of the sections Mathematical Physics and Algebraic
Geometry celebrates a historic period of more than two decades of
remarkably fruitful interactions between physics and mathematics. The
`unreasonable effectiveness,' depth and universality of quantum field
theory ideas in mathematics continue to amaze, with applications not only
to algebraic geometry, but also to topology, global analysis,
representation theory, and many more fields. The impact of string theory
has been particularly striking, leading to such wonderful developments
as mirror symmetry, quantum cohomology, Gromov-Witten theory, invariants
of three-manifolds and knots, all of which were discussed at length at
previous Congresses.

Many of these developments find their origin in two-dimensional
conformal field theory (CFT) or, in physical terms, in the
first-quantized, perturbative formulation of string theory. This is
essentially the study of sigma models or maps of Riemann surfaces
$\Sigma$ into a space-time manifold $X$. Through the path-integral over
all such maps a CFT determines a partition function
$Z_g$ on the moduli space $\cM_g$ of genus $g$ Riemann surfaces.
String amplitudes are functions $Z(\l)$, with $\l$ the string coupling
constant, that have asymptotic series of the form
\be
Z(\l) \sim \sum_{g \geq 0} \l^{2g-2} \int_{\cM_g} Z_g.
\ee

But string theory is more than a theory of Riemann surfaces. Recently it
has become possible to go beyond perturbation theory through conceptual
breakthroughs such as string duality \cite{witten-duality} and D-branes
\cite{polchinski}. Duality transformations can interchange the string
coupling $\l$ with the much better understood geometric moduli of the
target space $X$. D-Branes are higher-dimensional extended objects that
give rise to special cycles $Y \subset X$ on which the Riemann surface
can end, effectively leading to a {\it relative} form of string theory. 

One of the most important properties of branes is that they can have
multiplicities. If $k$ branes coincide a non-abelian $U(k)$ gauge
symmetry appears. Their `world-volumes' carry Yang-Mills-like quantum
field theories that are the analogues of the two-dimensional CFT on the
string world-sheet. The geometric realization as special cycles (related
to the theory of calibrations) has proven to be a powerful tool to
analyze the physics of these field theories. The mathematical
implications are just starting to be explored and hint at an intricate
generalization of the CFT program to higher dimensions.

This lecture is a review of work done on one of these non-perturbative
objects, the fivebrane, over the past years in collaboration with Erik
Verlinde, Herman Verlinde and Gregory Moore
\cite{5brane,dyon,ell-genus,5d}. I thank them for very enjoyable and
inspiring discussions.

\section{Fivebranes}

One of the richest and enigmatic objects in non-perturbative string
theory is the so-called fivebrane, that can be considered as a
six-dimensional cycle $Y$ in space-time. Dimension six is special since,
just as in two dimensions, the Hodge star satisfies $*^2=-1$ and one can
define {\it chiral} or `holomorphic' theories. The analogue of a free
chiral field theory is a 2-form `connection' $B$ with a self-dual
curvature $H$ that is locally given as $H=dB$ but that can have a `first
Chern class' $[H/2\pi] \in H^3(Y,\Z)$. (Technically it is a Deligne
cohomology class, and instead of a line bundle with connection it
describes a 2-gerbe on $Y$.) A system of $k$ coinciding fivebranes is
described by a 6-dimensional conformal field theory, that is morally a
$U(k)$ non-abelian 2-form theory. Such a theory is not known to exist at
the classical level of field equations, so probably only makes sense as
a quantum field theory.

One theme that we will not further explore here is that (at least for
$k=1$) the fivebrane partition function $Z_Y$ can be obtained by
quantizing the intermediate Jacobian of $Y$, very much in analogy with
the construction of conformal blocks by geometric quantization of the
Jacobian or moduli space of vector bundles of a Riemann surface
\cite{witten-M5}. This leads to interesting relations with the geometry
of moduli spaces of Calabi-Yau three-folds and topological string
theory. In fact there is even a seven-dimensional analogue of Chern-Simons
theory at play.
 
The fivebrane theory is best understood on manifolds of the product form
$Y=X\times T^2$, with $X$ a 4-manifold. In the limit where the volume
of the two-torus goes to zero, it gives
a $U(k)$ Yang-Mills theory as studied in \cite{vafa-witten}. In that
case the partition function computes the Euler number of the moduli
space of $U(k)$ instantons on $X$. In the $k=1$ case this relation follows
from the decomposition of the 3-form 
\be
H = F_+ \wedge dz + F_- \wedge d\zbar
\ee
with $F_{\pm}$ (anti)-self-dual 2-forms on $X$. In this way holomorphic
fields on $T^2$ are coupled to self-dual instantons on $X$. The obvious
action of $SL(2,\Z)$ on $T^2$ translates in a deep quantum symmetry
($S$-duality) of the 4-dimensional Yang-Mills theory.

Actually, the full fivebrane theory is much richer than a 6-dimensional
CFT. It is believed to be a six-dimensional {\it string} theory that
does not contain gravity and that reduces to the CFT in the
infinite-volume limit. We understand very little about this new class of
string theories, other than that they can be described in certain limits
as sigma models on instanton moduli space
\cite{strominger-vafa,5d,abkss,witten-higgs}. As we will see, this
partial description is good enough to compute certain topological
indices, where only so-called BPS states contribute.

\section{Conformal field theory and modular forms}

One of the striking properties of conformal field theory is the natural
explanation it offers for the modular properties of the characters of
certain infinite-dimensional Lie algebras such as affine Kac-Moody
algebras. At the hart of this explanation---and in fact of much of the
applications of quantum field theory to mathematics---lies the
equivalence between the Hamiltonian and Lagrangian formulation of
quantum mechanics \cite{witten-icm}. For the moment we consider a {\it
holomorphic} or chiral CFT.

In the Hamiltonian formulation the partition function on an elliptic
curve $T^2$ with modulus $\t$ is given by a trace over the Hilbert space
$\cH$ obtained by quantization on $S^1 \times \R$. For a sigma model with target space $X$, this Hilbert space
will typically consist of $L^2$-functions on the loop space $\cL
X$. It forms a representation of the algebra of quantum observables
and is $\Z$-graded by the momentum operator $P$ that generates the
rotations of $S^1$.  For a chiral theory $P$ equals the holomorphic
Hamiltonian $L_0=z\partial_z$. The character of the representation is
then defined as
\be
Z(\t)= \Tr _{\strut \! \cH} q^{P-{c\over 24}}
\ee
with $q=e^{2\pi i \tau}$ and $c$ the central charge of the Virasoro
algebra.  The claim is that this character is always 
a suitable modular form
for $SL(2,\Z)$, \ie, it transforms covariantly under linear fractional
transformations of the modulus $\t$.

In the Lagrangian formulation $Z(\t)$ is computed from the path-integral
over maps from $T^2$ into $X$. The torus $T^2$ is obtained by gluing
the two ends of the cylinder $S^1 \times \R$, which is the geometric
equivalent of taking the trace.
$$
\Tr\ \insertfig{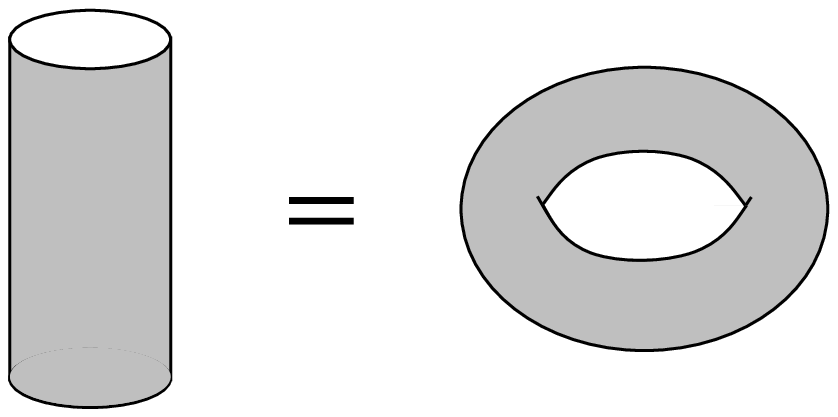}{3}
$$ 
Modularity is therefore built in from the start, since $SL(2,\Z)$ is the
`classical' automorphism group of the torus $T^2$.

The simplest example of a CFT consists of $c$ free chiral scalar fields
$x: \Sigma\to V\cong \R^c$. Ignoring the zero-modes, the chiral operator
algebra is then given by an infinite-dimensional Heisenberg algebra that
is represented on the graded Fock space
\be
\cH_q = \bigotimes_{n>0} S_{q^n} V.
\label{fock}
\ee
Here we use a standard notation for formal sums of (graded) symmetric 
products
\be
S_qV = \bigoplus_{N \geq 0} q^N S^N V,\qquad 
S^N\!V=V^{\otimes N}/S_N.
\ee
The partition function is then evaluated as
\be
Z(\t) = q^{-{c\over 24}} \prod_{n>0} (1-q^n)^{-c} = \eta(q)^{-c}
\ee
and is indeed a modular form of $SL(2,\Z)$ of weight $-c/2$ (with
multipliers if $c \not\equiv 0$ mod 24.) The `automorphic correction'
$q^{-c/24}$ is interpreted as a regularized sum of zero-point
energies that naturally appear in canonical quantization.

\section{String theories and automorphic forms}

The partition function of a {\it string} theory on a manifold $Y$ will have
automorphic properties under a larger symmetry group that reflects the
`stringy' geometry of $Y$. For example, if we choose $Y= X \times S^1 \times
\R$, with $X$ compact and simply-connected, quantization will lead to a
Hilbert space $\cH$ with a natural $\Z\oplus\Z$ gradation. Apart from
the momentum $P$ we now also have a winding number $W$ that labels the
components of the loop space $\cL Y$. Thus we can define a two-parameter
character
\be
Z(\s,\t) = \Tr_{\strut \! \cH} \bigl(p^W \! q^{P}\bigr),
\label{char}
\ee
with $p= e^{2\pi i \s}, q=e^{2\pi i \tau}$, with both $\s,\t$ in the
upper half-plane $\H$. We claim that $Z(\s,\t)$ is typically the
character of a generalized Kac-Moody algebra \cite{GKM} and an
automorphic form for the arithmetic group $SO(2,2;\Z)$.

The automorphic properties of such characters become evident by changing
again to a Lagrangian point of view and computing the partition function
on the compact manifold $X \times T^2$. The $T$-duality or `stringy' 
symmetry group of $T^2$ is 
\be
SO(2,2;\Z) \cong PSL(2,\Z) \times PSL(2,\Z) \ltimes \Z_2,
\ee
where the two $PSL(2,\Z)$ factors act on $(\s,\t)$ by separate
fractional linear transformations and the mirror map $\Z_2$ interchanges
the complex structure $\t$ with the complexified K\"ahler class $\s\in
H^2(T^2,\C)$. This group appears because a string moving on $T^2$ has
both a winding number $w\in \L=H_1(T^2;\Z)$ and a momentum vector
$p\in\L^*$. The 4-vector $k=(w,p)$ takes value in the even, self-dual
Narain lattice $\G^{2,2}=\L\oplus \L^*$ of signature $(2,2)$ with
quadratic form $k^2=2w\cdot p$ and automorphism group $SO(2,2;\Z)$.

In the particular example we will discuss in detail in the next
sections, where the manifold $X$ is a Calabi-Yau space, there will be
an extra $\Z$-valued quantum number and the Narain lattice will be
enlarged to a signature $(3,2)$ lattice. Correspondingly, the
automorphic group will be given by $SO(3,2,\Z) \cong Sp(4,\Z)$.

\section{Quantum mechanics on the Hilbert scheme}

As we sketched in the introduction, in an appropriate gauge the
quantization of fivebranes is equivalent to the sigma model (or quantum
cohomology) of the moduli space of instantons. More precisely,
quantization on the six-manifold $X\times S^1 \times \R$, gives a graded
Hilbert space
\be 
\cH_p = \bigoplus_{N\geq 0} p^N \cH_N ,
\ee 
where $\cH_N$ is the Hilbert space of the two-dimensional
supersymmetric sigma model on the moduli space of $U(k)$ instantons of
instanton number $N$ on $X$. If $X$ is an algebraic complex surface,
one can instead consider the moduli space of stable vector bundles of
rank $k$ and $ch_2=N$. This moduli space can be compactified by
considering all torsion-free coherent sheaves up to equivalence. In
the rank one case it coincides with the Hilbert scheme of points on $X$.
This is a smooth resolution of the symmetric product $\SNX$. (We note
that for the important Calabi-Yau cases of a $K3$ or abelian
surface the moduli spaces are all expected to be hyper-K\"ahler
deformations of $S^{Nk}\!X$.)

The simplest type of partition function will correspond to the Witten index. For
this computation it turns out we can replace the Hilbert scheme by the
more tractable orbifold $S^N\!X$. For a smooth manifold $M$ the Witten
index computes the superdimension of the graded space of ground states
or harmonic forms, which is isomorphic to $H^*(M)$, and therefore equals
the Euler number $\chi(M)$. 

For an orbifold $M/G$ the appropriate generalization is the {\it
orbifold} Euler number. If we denote the fixed point locus of $g\in G$ as
$M^g$ and centralizer subgroups as $C_g$, this is defined as a sum over
the conjugacy classes $[g]$ 
\be
\chi_{orb}(M/G) = \sum_{[g]} \chi_{top}(M^g/C_g).
\ee
For the case of the symmetric product $S^N\!X$ this expression can be
straightforwardly computed, as we will see in the next section, 
and one finds

\medskip

{\bf Theorem 1} \cite{hirzebruch} --- {\sl The orbifold Euler numbers of
the symmetric products $S^N\! X$ are given by the generating function}
$$
\chi_{orb}(S_p X)= \prod_{n>0} (1-p^n)^{-\chi(X)}.
$$

Quite remarkable, if we write $p=e^{2\pi i \t}$, the formal sum of Euler
numbers is (almost) a modular form for $SL(2,\Z)$ of weight
$\chi(X)/2$. This is in accordance with the interpretation as a
partition function on $X \times T^2$ and the $S$-duality of the
corresponding Yang-Mills theory on $X$ \cite{vafa-witten}. 

A much deeper result of G\"ottsche tells us that the same result holds
for the Hilbert scheme \cite{goettsche}. In fact, in both cases one can 
also compute the full cohomology and express it as the Fock space,
generated by an  infinite series of copies of $H^*(X)$ shifted in
degree \cite{gs}
\be
H^*(S_pX) \cong \bigotimes_{n>0} S_{p^n} H^{*-2n+2}(X).
\ee
Comparing with (\ref{fock}) we conclude that the Hilbert space of ground
states of the fivebrane is the Fock space of a chiral CFT. This does not
come as a surprise given the remarks in the introduction. One can also
derive the action of the corresponding Heisenberg algebra using
correspondences on the Hilbert scheme \cite{nakajima}.

\section{The elliptic genus}

We now turn from particles to strings. To compute the fivebrane
string partition function on $X\times T^2$, we will have to study
the two-dimensional supersymmetric sigma model on the moduli space of
instantons on $X$. Instead of the full partition function we will
compute again a topological index --- the elliptic genus. Let us briefly
recall its definition.

For the moment let $X$ be a general complex manifold of dimension $d$.
Physically, the elliptic genus is defined as the partition function of
the corresponding $N=2$ supersymmetric sigma model on a torus with
modulus $\t$ \cite{ell}
\be
\chi(X;q,y) = \Tr_{\strut \! \cH}\Bigl(
(-1)^{F_L+F_R}  y^{F_L} q^{L_0 -{d\over 8}}\Bigr),
\ee
with $q = e^{2\pi i \t}$, $y = e^{2\pi i z}$, $z$ a point on $T^2$. Here
$\cH$ is the Hilbert space obtained by quantizing the loop space $\cL X$
(formally the space of half-infinite dimensional differential forms).
The Fermi numbers $F_{L,R}$ represent (up to an infinite shift that is
naturally regularized) the bidegrees of the Dolbeault differential forms
representing the states. The elliptic genus counts the number of string
states with $\Lbar_0=0$. In terms of topological sigma models, these
states are the cohomology classes of the right-moving BRST operator
$Q_R$. In fact, if we work modulo $Q_R$, the CFT gives a {\it
cohomological} vertex operator algebra.

Mathematically, the elliptic genus can be understood as the
$S^1$-equivariant Hirzebruch $\chi_y$-genus of the loop space of
$X$. If $X$ is Calabi-Yau the elliptic genus has nice
modular properties under $SL(2,\Z)$. It is a weak Jacobi form
of weight zero and index $d/2$ (possibly with multipliers). The coefficients
in its Fourier expansion
\be
\chi(X;q,y) = \sum_{m\geq 0,\;\ell} c(m,\ell) q^my^\ell
\ee
are integers and can be interpreted as indices of
twisted Dirac operators on $X$. For a $K3$
surface one finds the unique (up to scalars) Jacobi form of weight zero
and index one,
that can expressed in elementary theta-functions as
\be
\chi(K3;q,y) = 2^3 \cdot \sum_{{\rm even}\ \a} {\vartheta^2_\a(z;\t)
/ \vartheta^2_a(0;\t)}.
\ee

\section{Elliptic genera of symmetric products}

We now want to compute the elliptic genus of the moduli spaces of vector
bundles, in particular of the Hilbert scheme. Again, we first turn to
the much simpler symmetric product orbifold $\SNX$. 

The Hilbert space of a two-dimensional sigma model on any orbifold $M/G$
decomposes in sectors labeled by the conjugacy classes $[g]$ of $G$,
since $\cL(M/G)$ has disconnected components of twisted loops satisfying
\be
x(\s + 2\pi) = g \cdot x(\s), \qquad g\in G.
\ee
In the case of the symmetric product orbifold $X^N\!/S_N$ these twisted
sectors have an elegant interpretation \cite{ell-genus}. The
conjugacy classes of the symmetric group $S_N$ are labeled by partitions of
$N$,
\be
[g] = (n_1) \cdots (n_s),\qquad \sum_i n_i = N,
\ee
where $(n_i)$ denotes an elementary cycle of length $n_i$. A loop on $\SNX$
satisfying this twisted boundary condition can therefore be visualized as
$$
\insertfig{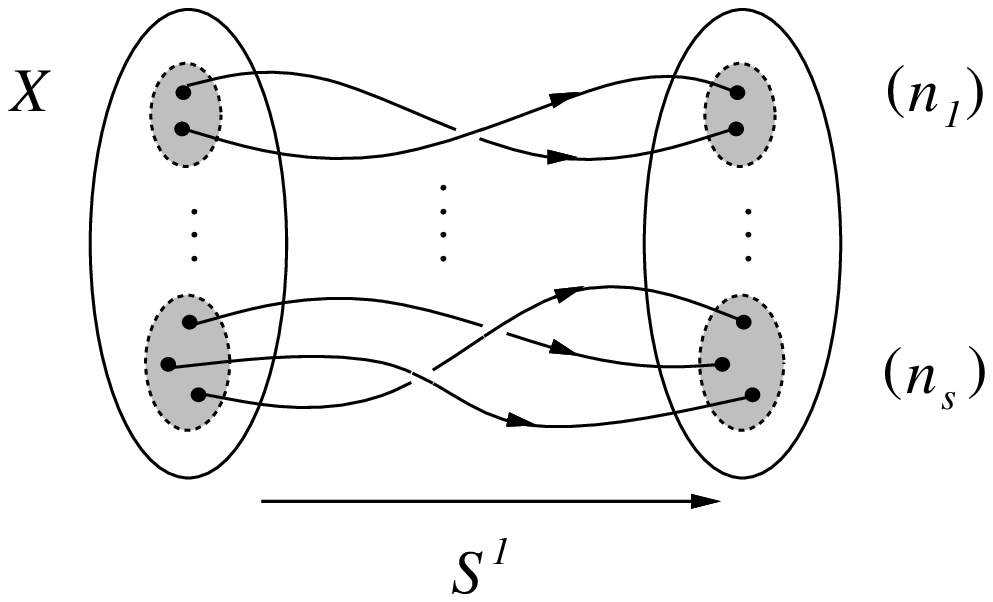}{6.1}
$$ 
As is clear from this picture, one loop on $\SNX$ is {\it not}
necessarily describing $N$ loops on $X$, but instead can describe $s\leq
N$ loops of length $n_1,\ldots,n_s$. By length $n$ we understand that
the loop only closes after $n$ periods. Equivalently, the action of the
canonical circle action is rescaled by a factor $1/n$. 

In this way we obtain a `gas' of strings labeled by the additional
quantum number $n$. The Hilbert space of the formal sum $S_pX$ can 
therefore be written as
\be 
\cH(S_pX) = \bigotimes_{n>0} S_{p^n} \cH_n(X) .
\ee
Here $\cH_n(X)$ is the Hilbert space obtained by quantizing a single
string of length $n$. It is isomorphic to the subspace $P \equiv 0$ (mod $n$)
of the single string Hilbert space $\cH(X). $
>From this result one derives

\medskip

{\bf Theorem 2} \cite{ell-genus} --- {\sl Let $X$ be a Calabi-Yau
manifold, then the orbifold elliptic genera of the symmetric products
$\SNX$ are given by the generating function}
$$
\chi_{orb}(S_pX;q,y) = 
\prod_{n>0,\,m\geq 0,\,\ell} (1-p^nq^m y^\ell)^{-c(nm,\ell)}.
$$

In the limit $q \to 0 $ the elliptic genus reduces to the Euler number
and we obtain the results from \S5. Only the constant loops survive and,
since twisted loops then localize to fixed point sets, we recover the
orbifold Euler character prescription and Theorem 1.

\section{Automorphic forms and generalized Kac-Moody algebras}

The fivebrane string partition function is obtained from the above
elliptic genus by including certain `automorphic corrections' and is
closely related to an expression of the type studied extensively by Borcherds
\cite{borcherds} with the infinite product representation\footnote{Here
the positivity condition means: $n,m\geq0$, and $\ell>0$ if $n=m=0$. The
`Weyl vector' $(a,b,c)$ is defined by $a=b= \chi(X)/24,$ and
$c= -{1\over 4}\sum_\ell |\ell|c(0,\ell)$.}
\be
\Phi(\s,\t,z) = p^a q^b y^c \prod_{(n,m,\ell)>0}(1-p^n q^m 
y^\ell)^{c(nm,\ell)}
\ee
For general Calabi-Yau space $X$ it can be shown, using
the path-integral representation, that the
product $\Phi$ is an automorphic form of weight $c(0,0)/2$ for the group
$SO(3,2,\Z)$ for a suitable quadratic form of signature $(3,2)$
\cite{harvey-moore,kawai,neumann}. 

In the important case of a $K3$ surface $\F$ is the square of 
a famous cusp form of $Sp(4,\Z) \cong SO(3,2,\Z)$ of weight 10, 
\be
\F(\s,\t,z) = 2^{-12} \prod_{even\ \a} \vartheta[\a](\W)^2
\ee
the product of all even theta-functions on a genus-two surface $\Sigma $
with period matrix
\be
\W = \twomatrix \s z z \t, \qquad \det{\rm Im}\,\W > 0.
\ee
Note that $\F$ is the 12-th power of the holomorphic determinant of the
scalar Laplacian on $\Sigma$, just as $\eta^{24}$ is on an elliptic
curve. The quantum mechanics limit $\s\to i\infty$ can be seen as the
degeneration of $\Sigma$ into a elliptic curve.

In the work of Gritsenko and Nikulin \cite{gritsenko} it is shown that
$\F$ has an interpretation as the denominator of a generalized Kac-Moody
algebra. This GKM algebra is constructed out of the cohomological vertex
algebra of $X$ similar as in the work of Borcherds. This algebra of BPS
states is induced by the string interaction, and should also have an
algebraic reformulation in terms of correspondences as in
\cite{harvey-moore}.

\section{String interactions}

Usually in quantum field theory one first quantizes a single particle on
a space $X$ and obtains a Hilbert space $\cH=L^2(X)$. Second
quantization then corresponds to taking the free symmetric algebra
$\bigoplus_N S^N\cH$. Here we effective reversed the order of the two
operations: we considered quantum mechanics on the `second-quantized'
manifold $\SNX$. (Note that the two operations do not commute.) In this
framework it is possible to introduce interactions by deforming the
manifold $\SNX$, for example by considering the Hilbert scheme or the
instanton moduli space. It is interesting to note that there is another
deformation possible.

To be concrete, let $X$ be again a $K3$ surface. Then $\SNX$ or $\HNX$ is an
Calabi-Yau of complex dimension $2N$. Its moduli space is
unobstructed and $21$ dimensional --- the usual 20 moduli of the $K3$ 
surface plus one extra modulus. This follows essentially from
\be
h^{1,1}(\SNX)=h^{1,1}(X) + h^{0,0}(X).
\ee
The extra cohomology class is dual to the small diagonal, where two
points coincide, and the corresponding modulus controls the blow-up of
this $\Z_2$ singularity. Physically it is represented by a $\Z_2$ twist
field that has a beautiful interpretation, that mirrors a construction
for the 10-dimensional superstring \cite{matrix} --- it describes the
joining and splitting of strings. Therefore the extra modulus can be
interpreted as the string coupling constant $\l$ \cite{5d,witten-higgs}.

The geometric picture is the following. Consider the sigma model with
target space $\SNX$ on the world-sheet $\P^1$. A map $\P^1 \to \SNX$ can
be interpreted as a map of the $N$-fold {\it unramified} cover of $\P^1$
into $X$. If we include the deformation $\l$ the partition function has
an expansion
\be
Z(\l) \sim \sum_{n\geq 0} \l^n Z_n,
\ee
where $Z_n$ is obtained by integrating over maps with $n$ simple branch
points. In this way higher genus surfaces appear as non-trivial $N$-fold
branched covers of $\P^1$. The string coupling has been given a
geometric interpretation as a modulus of the Calabi-Yau $\SNX$.

It is interesting to note that this deformation has an alternative
interpretation in terms of the moduli space of instantons, at least on
$\R^4$. The deformed manifold with $\l \not= 0$ can be considered as the
moduli space of instantons on a {\it non-commutative} version of $\R^4$
\cite{nekrasov-schwarz}.

%%--------------------Here the manuscript ends--------------------------------
\Addresses
\end{document}